\documentstyle[aps,twocolumn]{revtex}
\def\frac#1#2{{#1\over#2}}

\def\sqr#1#2{{\vcenter{\hrule height.#2pt
 \hbox{\vrule width.#2pt height#1pt \kern#1pt
  \vrule width.#2pt}
 \hrule height.#2pt}}}
\mathchardef\bigbull='21017
\mathchardef\bigcirc='21016

\def\Dagcomp{^{\vphantom{\dagger}}}

\begin{document}
\ifpreprintsty\else
\twocolumn[\hsize\textwidth%
\columnwidth\hsize\csname@twocolumnfalse\endcsname
\fi
\title{\Large{\bf Luminescence Spectra of a Quantum-Dot Cascade Laser}}
\author{\large\bf Vadim Apalkov and Tapash Chakraborty}
\address{Max-Planck-Institut f\"ur Physik Komplexer Systeme, 
01187 Dresden, Germany}
\maketitle
\begin{abstract}
{\it A quantum cascade laser where the quantum wells in the active 
regions are replaced by quantum dots with their atom-like discrete 
energy levels is an interesting system to study novel features in 
optical spectroscopy. We study structures suitable for diagonal lasing 
transitions in coupled dots, and vertical lasing transitions in a single 
dot, in the active regions of the laser device. The luminescence spectra
as a function of electron number and dot size show that for diagonal
transitions, a significant amount of blue-shift in the emission 
spectra can be achieved by increasing electron population in the 
quantum dots as well as by decreasing the size of the dots. 
}
\end{abstract}
\ifpreprintsty\clearpage\else\vskip1pc]\fi
\narrowtext

Ever since the original work on quantum cascade laser (QCL) 
by Faist et al. \cite{first,vertical} in 1994,
the unipolar semiconductor laser based on 
intersubband transitions in coupled quantum wells have undergone 
rapid developments. QCLs created in InGaAs/AlInAs systems have 
achieved record high power outputs in the mid-infrared range that 
has the potential for wide-ranging applications \cite{power,apply}. 
Applications, oftened mentioned in the literature, are in
environmental sensing, pollution monitoring, medical diagnostics, etc.
QCLs in other material systems \cite{type2,gaas,proceed} have also 
shown promise for improved performance.

In this work, we study the optical properties of quantum cascade
lasers where the quantum wells in the active regions are replaced by 
quantum dots (QD). The latter objects, popularly known as artificial 
atoms \cite{qdbook}, where the electron motion is quantized in all 
three spatial directions, have been receiving much attention. These
zero-dimensional quantum confined systems are useful for investigating
the fundamental concepts of nanostructures \cite{qdbook,tarucha} as well 
as for its application potentials. In recent years, there has been 
considerable progress in quantum-dot laser research \cite{qdrev}. 
Because of their discrete atom-like states, quantum-dot lasers are 
expected to have better performance than the quantum-well 
lasers \cite{sakaki,vahala}. Developments of self-organizing
growth techniques that allow formation of high-density of quantum 
dots with nanometer dimensions rapidly enhanced the development 
of QD-laser research, where the performance is now comparable to that 
of quantum-well lasers \cite{laser1,laser2}. Researchers have also 
found important applications of quantum dots in storage devices
\cite{finley,petroff} and fluorescence markers \cite{marker}.

Here we combine the properties of these two very interesting 
nanostructures, the QCLs and the QDs, to explore the luminescnece spectra 
of a quantum-dot cascade laser. There has been already some suggestions 
in the literature that quantum-dot cascade lasers will significantly 
reduce the threshold current density by eliminating single phonon 
decays \cite{box,dot}. This prediction was based on the fact that 
quantization of electron motion in the plane would greatly inhibit 
single phonon decay, provided the dots are sufficiently small. There 
is however no report in the literature (theoretical or experimental)
as yet, on the physical properties of a quantum-dot cascade laser. 
In this work, we have explored the luminescnece spectra of quantum 
cascade lasers, both for vertical as well as diagonal lasing 
transitions, for various values of the dot size and number of electrons 
in the quantum dots. One advantage of the quantum-dot cascade laser
for theoretical studies over the quantum-well cascade laser is
that, for few electrons in the QD, most of the physical properties 
can be calculated exactly, albeit numerically \cite{qdbook}.

The single-electron Hamiltonian for our system is
\begin{equation}
{\cal H}'=\frac{p_x^2}{2m^*}+\frac{p_y^2}{2m^*}+V_{\rm plane}(x,y)
+\frac{p_z^2}{2m^*}+V_{\rm conf}(z)
\label{hamil}
\end{equation}
where the confinement potential in the $z$-direction 
[Fig.~\protect\ref{device}] is
$$V_{\rm conf}(z)=-{\rm e}Fz + \left\{
\begin{array}{ll}
0 & \mbox{for wells} \\
U_0 & \mbox{for barriers}
\end{array}
  \right.
$$
with $F$ being the electric field in the $z$-direction. The confinement
potential in the $xy$-plane is
$$V_{\rm plane}(x,y)= \left\{
\begin{array}{ll}
0 & \mbox{$|x|<L/2$ and $|y|<L/2$} \\
U_0 & \mbox{otherwise.}
\end{array}
  \right.
$$
The eigenfunctions and eigenvalues of the single-electron Hamiltonian
[\protect\ref{hamil}] is obtained from
$$\psi_{nmk}=\varphi_{n,k}(x)\varphi_{m,k}(y)\chi_k(z)$$
where 
\begin{eqnarray}
\label{ntype}
\Bigg[\frac{p_x^2}{2m^*}&+&V_{\rm plane}(x)\Bigg]\varphi_{n,k}(x)
= E_n\varphi_{n,k}(x)\\
\label{ktype}
\Bigg[\frac{p_z^2}{2m^*}&+&V_{\rm conf}(z)\Bigg]\chi_k(z)
={\tilde E}_k\chi_k(z)\\
\nonumber
E_{nmk}&=&E_{n,k}+E_{m,k}+{\tilde E}_k.\\
\nonumber
\end{eqnarray}
Because of the band nonparabolicity the electron mass depends on 
the total energy $m^*(E_{nmk})$ \cite{notes}. In our calculations that 
follow, we consider only two subbands in the $z$-direction $(k=1,2)$ and 
for a given $k$ all possible states in the $xy$-plane with the condition, 
$E_{nmk}<U_0$.

The solutions of Eq.~(\protect\ref{ntype}) are usual $\cos\ (\sin)$-type
$$\varphi_{n,k}(x)= {\cal N}\left\{
\begin{array}{ll}
{{\scriptstyle\cos}\atop{\scriptstyle\sin}}\left(
\sqrt{2m\Dagcomp_wE_n}L/2\right) \\
\times{\rm e}^{-\sqrt{2m\Dagcomp_b(U_0-E_n)}(X-L/2)},& \mbox{$x>L/2$} \\
{{\Large\cos}\atop{\Large\sin}}\left(\sqrt{2m\Dagcomp_wE_n}x\right),
& \mbox{$|x|<L/2$}\\
{{\Large\cos}\atop{-\Large\sin}}\left(\sqrt{2m\Dagcomp_wE_n}L/2\right)\\
\times{\rm e}^{-\sqrt{2m\Dagcomp_b(U_0-E_n)}(X+L/2)},& \mbox{$x<-L/2$} \\
\end{array}
  \right.
$$
where $m\Dagcomp_w (m\Dagcomp_b)$ is the electron effective mass in the
well (barrier). The energy $E_n$ was calculated numerically from
$$\begin{array}{ll}
\tan\kappa=\sqrt{\frac{m\Dagcomp_w}{m\Dagcomp_b}}\sqrt{
\frac{\gamma^2-\kappa^2}{\kappa^2}}, & \mbox{for even solutions 
} \\
\tan\kappa=-\sqrt{\frac{m\Dagcomp_b}{m\Dagcomp_w}}\sqrt{
\frac{\kappa^2}{\gamma^2-\kappa^2}}, & \mbox{for odd solutions 
} \\
\end{array}
$$
where $\kappa^2=E_n\frac{L^2}2m\Dagcomp_w$, $\gamma^2=U_0\frac{L^2}2m
\Dagcomp_w$. Solutions of Eq.~(\protect\ref{ktype}) are obtained 
numerically for the two lowest states (subbands) shown in 
Fig.~\protect\ref{device}. Due to the $x\leftrightarrow y$ symmetry, 
some of the levels are two-fold degenerate (for example, 
$E_{122}=E_{212}$).

\begin{figure}
\begin{center}
\begin{picture}(120,130)
\put(0,0){\includegraphics{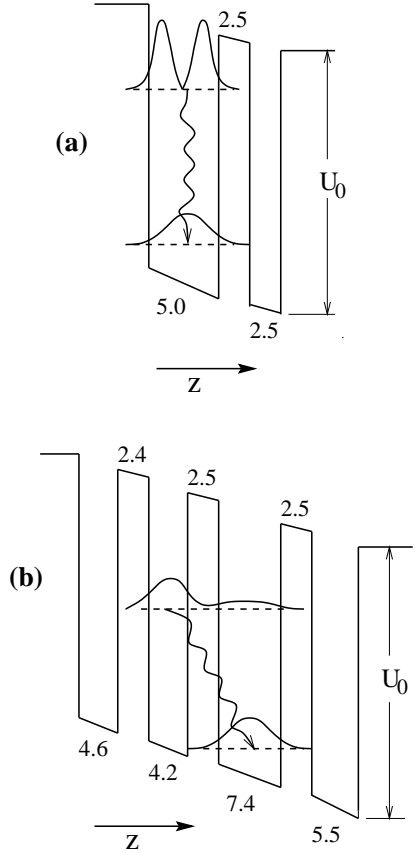}}
\end{picture}
\vspace*{7.0cm}
\caption{Energy band diagram (schematic) of the active region of
a quantum cascade laser structure and (a) vertical lasing transition
under an average applied electric field of 85 kV/cm and (b) diagonal
transition under a field of 55 kV/cm. The relevant wave functions 
(moduli squared) as well as the transition corresponding to the laser 
action are also shown schematically. The numbers (in nm) are the well 
(Ga$_{0.47}$In$_{0.53}$As) and barrier (Al$_{0.48}$In$_{0.52}$As) widths.
For vertical transitions, these parameters are taken from \protect\cite{box}.
Material parameters considered in this work are: electron
effective mass $m_e^*$ (Ga$_{0.47}$In$_{0.53}$As)=0.043 $m_0$,
$m_e^*$ (Al$_{0.48}$In$_{0.52}$As)=0.078 $m_0$, the conduction band
discontinuity, $U_0=520$ meV, the nonparabolicity coefficient,
$\gamma_w=1.3\times10^{-18}$ m$^2$ for the well and
$\gamma_b=0.39\times10^{-18}$ m$^2$ for the barrier. 
}
\label{device}
\end{center}
\end{figure}

>From the single-electron basis functions, we construct the $N$--electron 
basis
\begin{eqnarray}
\nonumber
&&\Phi_{n_im_ik_i}({\bf r}\Dagcomp_1,\cdots,{\bf r}\Dagcomp_N)\\
&&={\cal A}\,\psi_{n_1m_1k_1} ({\bf r}\Dagcomp_1)\sigma_1 \cdots 
\psi_{n\Dagcomp_Nm\Dagcomp_Nk\Dagcomp_N}({\bf r}\Dagcomp_N)
\sigma\Dagcomp_N 
\label{basis}
\end{eqnarray}
where, as usual, $\sigma_i$ is the spin part of the wave function
$[\sigma_i={1\choose0}$ or $\left(0\atop1\right)$] and
$\cal A$ is the antisymmetrization operator. The total many-electron 
Hamiltonian is written as
\begin{equation}
{\cal H}=\sum_{i=1}^N{\cal H}'_i+\sum_{i<j}^NV(|{\bf r}_i-{\bf r}_j|)
\label{total}
\end{equation}
where ${\cal H}'$ is given by Eq.~(\protect\ref{hamil}). For inter-electron
interactions we consider the Coulomb interaction
\begin{equation}
V(|{\bf r}_i-{\bf r}_j|)=\frac{e^2}{\epsilon|{\bf r}_i-{\bf r}_j|}.
\label{coulomb}
\end{equation}
The Hamiltonian matrix is then calculated in the basis 
(\protect\ref{basis}). The single-electron Hamiltonian has only diagonal 
contribution while the interaction term gives non-diagonal contributions
\begin{eqnarray*}
&&\langle{n_i}\Dagcomp_1{m_i}\Dagcomp_1{k_i}\Dagcomp_1;
{n_i}\Dagcomp_2{m_i}\Dagcomp_2{k_i}\Dagcomp_2|V|
{n_j}\Dagcomp_1{m_j}\Dagcomp_1{k_j}\Dagcomp_1;
{n_j}\Dagcomp_2{m_j}\Dagcomp_2{k_j}\Dagcomp_2\rangle\\
&&={\rm Tr}_\sigma\int\, d{\bf r}\Dagcomp_1 d{\bf r}\Dagcomp_2\,{\cal
A}\left[\Psi^*_{{n_i}\Dagcomp_1{m_i}\Dagcomp_1{k_i}\Dagcomp_1}
({\bf r}\Dagcomp_1)\sigma_{i_1}
\Psi^*_{{n_i}\Dagcomp_2{m_i}\Dagcomp_2{k_i}\Dagcomp_2}
({\bf r}\Dagcomp_2)\sigma_{i_2}\right] \\
&&\times V(|{\bf r}\Dagcomp_1-{\bf r}\Dagcomp_2|)\, {\cal A}\left[
\Psi^*_{{n_j}\Dagcomp_1{m_j}\Dagcomp_1{k_j}\Dagcomp_1}
({\bf r}\Dagcomp_2)\sigma_{j_1}
\Psi^*_{{n_j}\Dagcomp_2{m_j}\Dagcomp_2{k_j}\Dagcomp_2}
({\bf r}\Dagcomp_1)\sigma_{j_2}\right].\\
\end{eqnarray*}
The eigenvalues and eigenfunctions were calculated by exact (numerical)
diagonalization of the Hamiltonian matrix. As the many-body Hamiltonian
also has the $x\leftrightarrow y$ symmetry some of the states are
two-fold degenerate.

\begin{figure}
\begin{center}
\begin{picture}(120,130)
\put(0,0){\includegraphics{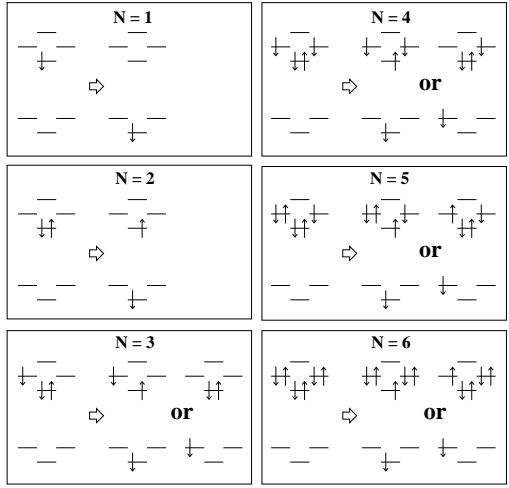}}
\end{picture}
\vspace*{2.5cm}
\caption{Energy levels and spin configurations and the allowed optical
transitions (schematic) for quantum dots with noninteracting electrons.
Here, $\uparrow$ denotes electrons with spin up and $\downarrow$ that
for spin down.}
\label{noninter}
\end{center}
\end{figure}

In the initial state (before optical emission) all the electrons are in
the second subband, $k=2$. In the final state (after optical emission)
one electron is in the first subband, $k=1$, and all other electrons are in
the second subband, $k=2$. The intensity of optical transitions is found
from the expression
\begin{eqnarray*}
&&{\cal I}_{if}(\omega)=\frac1Z\sum_{if}\delta(\omega-E_i+E_f)\\
&&\Bigg\vert\int\,\chi\Dagcomp_1(z)\,z\,
\chi\Dagcomp_2(z)\,dz\,\int \Phi_i^*(x\Dagcomp_1y\Dagcomp_1,\cdots,
x\Dagcomp_Ny\Dagcomp_N)\\
&&\times\Phi_f(x\Dagcomp_1y\Dagcomp_1,\cdots,x\Dagcomp_Ny\Dagcomp_N)\,
dx\Dagcomp_1 dy\Dagcomp_1\cdots dx\Dagcomp_Ndy\Dagcomp_N 
\Bigg\vert^2\\
&&\times\exp(-\beta E_i)\\
\end{eqnarray*}
where $Z=\sum_i\,{\rm e}^{-\beta E_i}$ is the partition function and
$\beta=1/kT$. In all our computation, we take $T=20$ K.

The energy levels and possible optical transitions for the noninteracting 
systems are sketched in Fig.~\ref{noninter}. Clearly,
there are two types of transitions for the noninteracting 
electron system: (a) from the one-particle ground state of the second 
subband to the one-particle ground state of the first 
subband and, (b) from the one-particle excited state of the
second subband to the one-particle excited state of the 
first subband. These are the transitions between states with 
same quantum number in the $xy$-plane. Energies of these 
transitions are different due to band nonparabolicity. The 
nonparabolicity can also allow transitions between states with 
different quantum numbers in the $xy$-plane but the 
intensity of these transitions are very small because the 
nonparabolicity has small  effect on the electron wave functions.
Therefore, we can safely assume that the optical transitions are 
allowed only between the 
states with the same quantum numbers in the $xy$-plane.
One also notices from Fig.~\ref{noninter} that optical transitions to 
the ground state of the final noninteracting system are forbidden for 
$N>2$. 

\begin{figure}
\begin{center}
\begin{picture}(120,130)
\put(0,0){\includegraphics{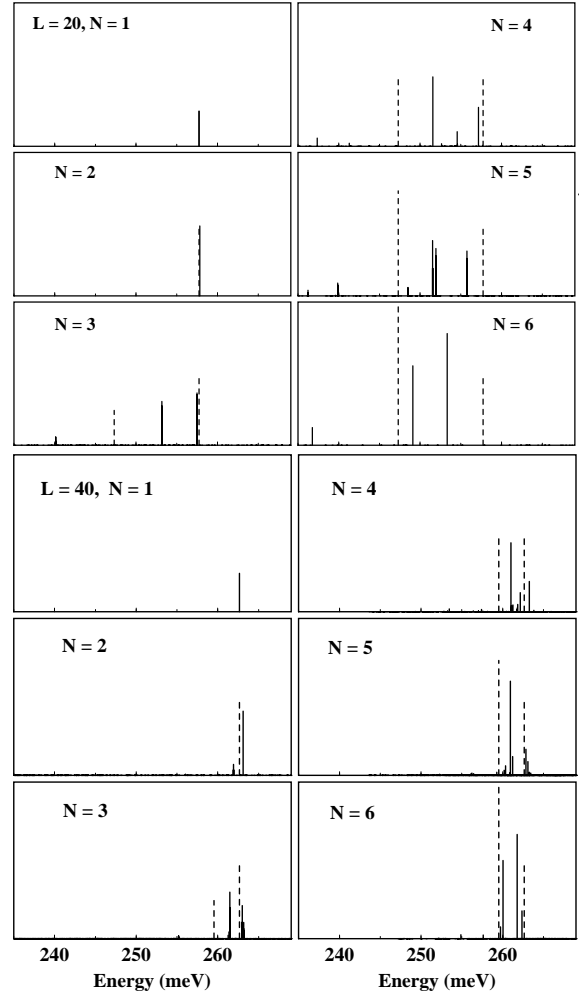}}
\end{picture}
\vspace*{9.5cm}
\caption{Luminescence spectra of a quantum-dot cascade laser and {\it
vertical} optical transitions for various values of the dot size 
($L$ in nm) and number of electrons $(N)$ in the dot. The dashed lines 
correspond to luminescence of noninteracting electron systems.
}
\label{vertical}
\end{center}
\end{figure}

In Fig.~\ref{vertical}, the optical spectra is shown for vertical 
transitions [Fig.~\ref{device}(a)] and for two sizes of quantum dots: 
$L=20$ nm and $L=40$ nm, and for different numbers of electrons in the
quantum dot. In all these cases the first moment of the emission spectra 
for the interacting system is almost the same as that for the noninteracting
system. This is bacause in the case of vertical transitions the 
Coulomb interaction between the electrons in the second subband and those
in the second and the first subbands are almost the same (electrons are 
localized in the same quantum well in the $z$-direction). The Coulomb 
interaction between electrons is about half the energy separation 
between the one-electron states in the $xy$-plane for $L=20$ nm and is of
the same order as the energy separation between $xy$ levels in $L=40$ nm.
That is why the interaction is more important for $L=40$ nm. 
For the two electron system there is  a small blue-shift 
of the emission line due to the interaction which increases with 
an increase of the size of the quantum well. In addition, there is also
a small red satellite at $L=40$ nm. For the six-electron system, the 
interaction results in redistribution of the intensities between peaks: 
the higher energy peak becomes more intense than that for the lower energy.
For 3,4 and 5 electrons in a noninteracting system, we have a degenerate 
initial state. The degeneracy is lifted by the interaction and for the 
four electon system the initial ground state is partially polarized as 
expected from Hund's rules. The interaction also results in the
apperance of satelites and at the same time the separation between 
the main peaks becomes smaller for the interacting system than for 
the noninteracting case.  

In Fig.~\ref{diagonal} the optical spectra is presented for diagonal 
optical transitions [Fig.~\ref{device} (b)] and for two sizes of quantum 
dots: $L=10$ nm and $L=20$ nm, and for different numbers of electrons.  
Interestingly, we notice the behavior characteristics of fully filled
shells for 2 and 6 electrons. The Coulomb 
interaction is about two times smaller then the separations between
shell levels in the $xy$-plane for $L=20$ nm and about six times smaller 
for $L=10$ nm. For the two-electron system, we have a single line 
for both non-interacting and interacting systems. For the six-electron 
system and $L=10$ nm, the emission spectra has the same two-peak structure 
as for the noninteraction system. For $L=10$ nm there is a small 
redistribution of intensities between the peaks while there is an 
additional line 
for $L=20$ nm. For 2,3 and 4 electrons in the dot, the interaction 
results in splitting of lines of the corresponding noninteracting 
systems. With increasing size of the quantum dots the lower 
energy lines become more intense. 
 
As the Coulomb interaction between electrons in the second subband 
is about two times larger then the Coulumb interaction between electrons 
in the second and in the first subband we have a highly non-symmetric 
system and as a result there is the {\it large blue shift} in all cases, 
compared to the noninteracting system. This blue shift decreases with 
increasing size of the quantum dots. Further, the blue shift increases 
with increasing number of electrons. For $L=10$ nm, there is a blue shift 
of the emission line of about 55 meV when the electron number is increased
from 1 to 6. 

\begin{figure}
\begin{center}
\begin{picture}(120,130)
\put(0,0){\includegraphics{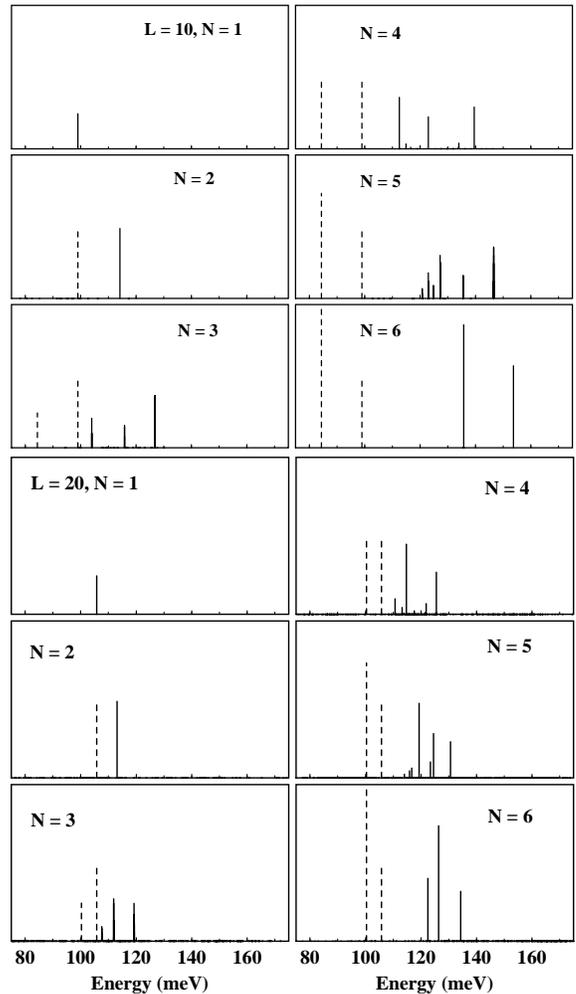}}
\end{picture}
\vspace*{9.5cm}
\caption{Luminescence spectra of a quantum-dot cascade laser and {\it
diagonal} optical transitions for various values of the dot size 
($L$ in nm) and number of electrons $(N)$ in the dot. The dashed lines 
correspond to luminescence of noninteracting electron systems.
}
\label{diagonal}
\end{center}
\end{figure}

In summary, we have studied the luminescence spectra of a quantum-dot
cascade laser suitable for vertical or diagonal transitions. The spectra
as a function of the dot size and electron numbers in the dot reflect
the atom-like character due to the presence of quantum dots. Most
interestingly, significant amount of blue-shift in the emission spectra 
can be achieved by increasing electron population in the quantum dots 
as well as by decreasing the size of the dots. This is most clearly
seen for the diagonal transitions. This opens up the possibility of tuning 
the laser emission frequency for diagonal transitions by changing the 
number of electrons in quantum dots and/or decreasing the size of
the dots.

We thank P. Fulde for his support and kind hospitality in Dresden.

\vfil

\begin{references}
\bibitem{notes}
In our calculations that follow, we consider only the main contribution
due to $\tilde E_k$, i.e., the mass depends only on $\tilde E_k$
(both in Eq.~[\protect\ref{ntype}] and [\protect\ref{ktype}]). This
approximation simplifies the analysis but will only result in small
changes in the final energies.
\end{references}
\end{document}